# Dimensional reduction at a quantum critical point

S. E. Sebastian[1], N. Harrison[2], C. D. Batista[3], L. Balicas[4], M. Jaime[2], P. A. Sharma[2], N. Kawashima[5] & I. R. Fisher[1]

[1]Geballe Laboratory for Advanced Materials and Department of Applied Physics, Stanford University, Stanford, California 94305, USA. [2]NHMFL, MS-E536, Los Alamos National Laboratory, Los Alamos, New Mexico 87545, USA. [3]Theoretical Division, Los Alamos National Laboratory, Los Alamos, New Mexico 87545, USA. [4]National High Magnetic Field Laboratory, Tallahassee, Florida 32310, USA. [5]Institute for Solid State Physics, University of Tokyo, Kashiwa, Chiba 277-8581, Japan.

**Competition between electronic ground states near a quantum critical point[1,2] (QCP)—the location of a zero-temperature phase transition driven solely by quantum-mechanical fluctuations—is expected to lead to unconventional behaviour in low-dimensional systems[3]. New electronic phases of matter have been predicted to occur in the vicinity of a QCP by two-dimensional theories [3–8], and explanations based on these ideas have been proposed for significant unsolved problems in condensed-matter physics, such as non-Fermi-liquid behaviour and high-temperature superconductivity. But the real materials to which these ideas have been applied are usually rendered three-dimensional by a finite electronic coupling between their component layers; a two-dimensional QCP has not been experimentally observed in any bulk three-dimensional system, and mechanisms for dimensional reduction have remained the subject of theoretical conjecture[9-11]. Here we show evidence that the Bose–Einstein condensate of spin triplets in the three-dimensional Mott insulator $BaCuSi_2O_6$ (refs 12–16) provides an experimentally verifiable example of dimensional reduction at a QCP. The interplay of correlations on a geometrically frustrated lattice causes the individual two-dimensional layers of spin-½ $Cu^{2+}$ pairs (spin dimers) to become decoupled at the QCP, giving rise to a two-dimensional QCP characterized by linear power law scaling distinctly different from that of its three-dimensional counterpart. Thus the very notion of dimensionality can be said to acquire an 'emergent' nature: although the individual particles move on a three-dimensional lattice, their collective behaviour occurs in lower-dimensional space.**

BaCuSi$_2$O$_6$ is a spin dimer system whose body-centred tetragonal crystal whose highly symmetric structure[17,18] (Fig. 1 inset) gives it unique advantages for tackling the fundamental role of dimensionality in the field of quantum criticality. The material consists of layers of spin dimers arranged vertically on an essentially square lattice in which exchange interactions are rotationally invariant around the crystalline $c$ axis—providing the ideal conditions for Bose–Einstein condensation. However, the dimer plaquettes are staggered between consecutive layers, as shown in Fig. 1 inset. This leads to geometrical frustration of the inter-layer antiferromagnetic interaction, $J_f$ (that is, the antiparallel arrangement of spins within and between layers conflict).

In zero magnetic field, each spin dimer has a singlet (spin $s$=0) ground state comprising $s$=½ Cu$^{2+}$ ions paired by an antiferromagnetic coupling constant $J$=4.45 meV (refs 12, 13, 19), with three degenerate triplet excited levels ($s$=1, $s_z$=1, 0, −1). Hence the ground state of the system is a quantum paramagnet consisting of a direct product of singlets, while the single triplet excitations become dispersive owing to the moderate intra-layer coupling $J'$=0.51 meV<<$J$ (refs 12, 13), but remain gapped. The effect of a magnetic field is to Zeeman-split the triplet bands in a linear fashion, closing the gap between the singlet level and the lowest energy $s_z$=1 triplet excitation (at wavevector $k_x$=$k_y$=π) at a critical magnetic field $H_{c1}$, which leads to an ordered state with a precisely field-tunable concentration (or density) $\rho \equiv m_z \equiv \langle s_z \rangle$ of triplets above $H_{c1}$ (here $m_z$ is the uniform magnetization and $\langle s_z \rangle$ is the expectation value of $s_z$).

Figure 1 shows the measured phase boundary separating the quantum paramagnetic phase from the magnetically ordered state. Ordering transitions are measured by torque magnetometry[13] in static magnetic fields of up to 28 T in the temperature range 30 mK to 1 K in a dilution refrigerator in the National High Magnetic Field Laboratory, Tallahassee. A finite torque is obtained by inclining $H$ at a small (<10°) angle to the crystalline $c$ axis, from which $m_Z \equiv \rho$ is extracted (see below).

We extract the critical scaling behaviour from the phase boundary in Fig.1 to determine the universality class of the QCP. Figure 2 shows the experimental value of the critical exponent $\nu$ determined from fitting points on the phase boundary in a sliding window. The exponent $\nu$ characterizes the power law[1,13] relating the ordering temperature to the proximity to the critical magnetic field $H_{c1}$: $T_c \propto (H-H_{c1})^\nu$. As the temperature is lowered, the system enters the region of universal behaviour (bright yellow shading). The value of $\nu$ tends to ⅔ in this region as $T$ tends to 0.5 K from above ($T_{win} \approx 0.9$ K), consistent with previous measurements[13]. However, below these temperatures there is a clear crossover to a value of $\nu = 1$ for the fitting range $T < 1$ K ($T_{win} < 0.65$ K). We show below that these exponents are characteristic of the three- and two-dimensional Bose–Einstein condensate (3D and 2D BEC) universality classes, respectively.

The system can be considered as a 3D Bose gas of interacting particles by neglecting the unoccupied higher energy $s_z = -1, 0$ triplet states (since $J_f$, $J' \ll J$), and replacing each dimer by an effective site that can be either empty (singlet state) or occupied by a hardcore boson ($s_z = 1$ triplet state)[12,20,21]. $H$ acts as a chemical potential, which at low magnetic fields is prohibitively high, and prevents population of the Bose gas, permitting population only above $H_{c1}$. The interacting bosons move on the frustrated lattice with kinetic energy provided by the $xy$-component of the inter-dimer Heisenberg interaction (which determines the effective mass), while the short range repulsion arises from the Ising or $z$-component. The kinetic energy term dominates in $BaCuSi_2O_6$, resulting in a BEC ordered state[12,22] characterized by a quantum-coherent superposition of the singlet and the $s_z=1$ triplet on all sites. This can be described as a canted XY-antiferromagnet in terms of the original spin degrees of freedom (shown in Fig. 1 inset), with the in-plane staggered magnetization $m_{xy}$ determining the amplitude and phase of the BEC order parameter $\langle b^\dagger \rangle$.

At the QCP ($H_{c1}$), the characteristic length of the amplitude fluctuations of $\langle b^\dagger \rangle$ diverges. Taking the scaling limit (lattice parameter $\to 0$), the original microscopic theory becomes an effective continuous field theory describing a dilute Bose gas[1,23]. In this field theory, the dynamical critical exponent $z = 2$, and hence the upper critical dimension of this QCP is $d_c = 2$. The correct critical exponents for $d \geq 2$ are therefore obtained using a mean-field theory, which leads to the following universal power laws for measurable quantities on approaching the QCP:

$\rho(T = 0) \equiv m_z(T = 0) \propto (H-H_{c1})$ \hfill (1a)

$m_z(H_{c1}) \propto T^{d/2}$ \hfill (1b)

$T_c \propto (H-H_{c1})^{2/d}$ \hfill (1c)

that is, the critical exponent $\nu \equiv 2/d$ for the BEC universality class.

According to equation (1c), the experimentally measured value of $\nu = \frac{2}{3}$ as $T$ tends to 0.5 K from above ($T_{win} \approx 0.9$ K, Fig. 2) is consistent with a 3D BEC[1,23–25], in agreement with previous measurements[13]. In contrast, the unexpected crossover to $\nu = 1$ in the fitting range $T < 1$ K ($T_{win} < 0.65$ K, Fig. 2) is a signature of 2D BEC behaviour. All experimental results for $T < 1$ K in Fig. 3a–c are consistent with the linear power law behaviours predicted by equation (1) for $d = 2$, thus demonstrating that the measured critical behaviour of the system in the temperature range 30 mK to 1 K belongs to the 2D BEC universality class. Ising-like order due to U(1) symmetry breaking terms can be ruled out, as this would lead to a critical exponent of 0.5 (ref. 1). The effect of disorder would be either a critical exponent satisfying the Harris criterion[24,26,27] ($\nu > 4/3$), or a smeared phase transition. Neither the experimentally measured exponent ($\nu = 1$), nor the sharp non-hysteretic transition in the magnetization characterized by the narrow peak in $d^2 m_z/dH^2$ which sharpens at low temperatures, are consistent with a disordered system.

To understand the origin of the dimensional reduction, we consider the effect of geometrical frustration as a consequence of the body-centred tetragonal lattice on which the Bose gas moves. The single particle dispersion relation comprises a term arising from the intra-layer exchange, $J'(\cos k_x + \cos k_y)$, and a term reflecting the inter-layer hopping, $2J_f \cos k_z \cos(k_x/2) \cos(k_y/2)$, as shown in Fig. 4a. At the dispersion minimum ($k_x = k_y = \pi$) where the bosons condense, the inter-layer hopping is always zero (that is, there is no dispersion along the $c$ axis). Equivalently, the phase of a single boson (with $k_x = k_y = \pi$) alternates cyclically on a plaquette of neighbouring lattice sites (shown in Fig. 4b), resulting in phase cancellation at each site of the adjacent layer. Hence the effect of geometrical frustration is to decouple adjacent layers, leading to a highly degenerate ground state, which can be described as an array of 2D BEC layers.

A competing effect arises from zero-point phase fluctuations, which generate an effective inter-layer coupling ($K$) and restore phase coherence along the $c$ axis[28,29]. However, because $K$ can equivalently be considered to be a consequence of pair tunnelling (as shown in Fig. 4c) and hence biquadratic in the order parameter, its strength decreases as $\rho^2$ for low densities of bosons (as recently shown using a spin-wave approach on the same spin lattice[28]). A measure of the '3D-ness' of the system is given by the size of the effective coupling $K$ relative to the temperature scale $k_B T_c$. As $T_c$ is proportional to $\rho$ in the proximity of the QCP (from equation (1) for $d = 2$), the ratio $K/k_B T_c$ (which is $\propto \rho$) is arbitrarily small close enough to the QCP, and inter-layer tunnelling is too small for the system to be observably 3D. By field tuning toward $H_{c1}$, we experimentally access this region, and hence observe critical power law behaviour consistent with the 2D BEC universality class. The energy scale of this 2D-critical behaviour is well separated from the very low temperatures at which weak longer range inter-layer interactions are anticipated to restore 3D behaviour; and hence 2D behaviour is observed over a significant range of temperature.

Away from the QCP, the fluctuation-induced inter-layer tunnelling increases rapidly with particle density, leading to distinctly 3D behaviour when $K(\rho)$ becomes comparable to $T_c$. The system may then be described by an effective unfrustrated model for the subsets of odd and even layers that assumes a constant effective inter-layer coupling, $J''$ (refs 12, 13). Tuning the system by means of the applied magnetic field from this unfrustrated region toward the QCP where geometrical frustration becomes effective therefore results in a marked crossover from 3D to 2D BEC power law behaviour (Fig. 2).

$BaCuSi_2O_6$ therefore provides a clear example of a system in which geometrical frustration causes the effective dimensionality to become reduced at the QCP, leading to 2D collective excitations despite the 3D nature of the system. Although inter-layer decoupling due to geometrical frustration features as a possible explanation for the puzzling experimental observation of non-Fermi-liquid behaviour at the QCP in body-centred tetragonal heavy fermion intermetallics[3-8], theoretical reasoning[28] has been used to assert that it is impossible to realize reduced dimensionality in any system by the mechanism of geometrical frustration. The experimentally observed dimensional reduction in $BaCuSi_2O_6$ provides a counter-example, and constitutes a proof of principle that dimensionality can become an emergent property of a QCP.

**Acknowledgements** N.H., C.D.B., M.J. and P.A.S. acknowledge Laboratory Directed Research and Development (LDRD) support at LANL. S.E.S and I.R.F. acknowledge National Science Foundation (NSF) support. Experiments performed at the NHMFL, Tallahassee were supported by the NSF, the State of Florida, and the Department of Energy. We thank T. P. Murphy, E. C. Palm, P. Tanedo and P. B. Brooks for experimental assistance, and acknowledge discussions with A. G. Green, E.-A. Kim, S. A. Kivelson, D. I. Santiago and J. Schmalian. I.R.F. acknowledges support from the Alfred P. Sloan Foundation and S.E.S. from the Mustard Seed Foundation.

**Author Information** Reprints and permissions information is available at npg.nature.com/reprintsandpermissions. The authors declare no competing financial interests. Correspondence and requests for materials should be addressed to S.E.S. (suchitra@stanfordalumni.org).

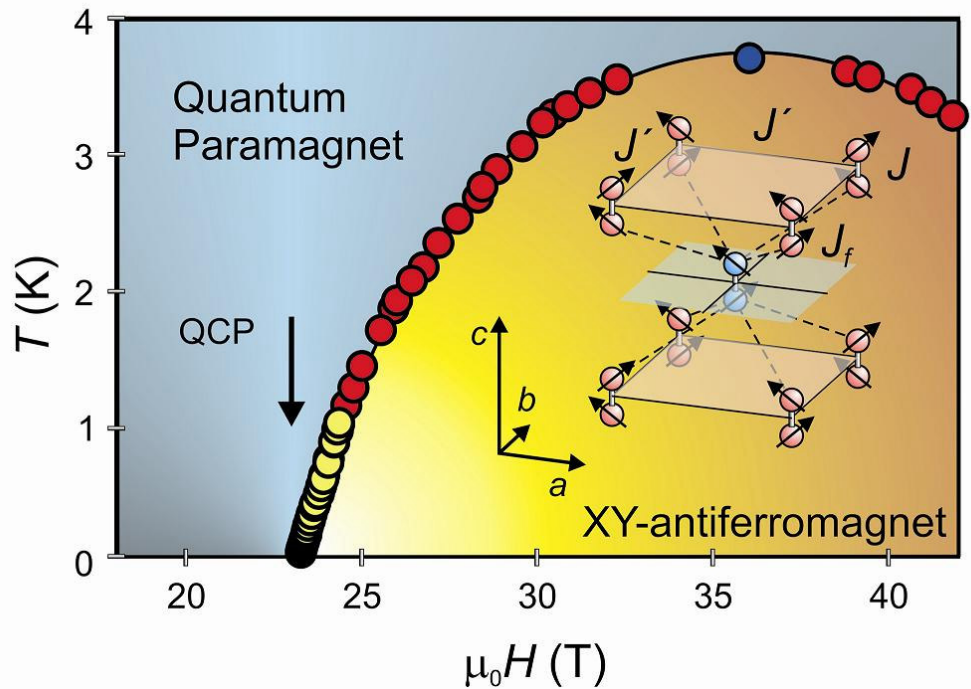

**Figure 1 | Experimentally obtained phase boundary of $BaCuSi_2O_6$.** The phase transition to the magnetically ordered state is shown over the entire temperature range (error bars are smaller than the symbol size); the shading varies radially from dark to light, reflecting the reduction in dimensionality near the QCP. Experimental data obtained from torque magnetometry in a dilution refrigerator are in yellow, previous experimental data from ref. 13 are in red (torque magnetometry and magnetocaloric effect in a $^3$He refrigerator) and blue (specific heat in a $^4$He cryostat). The inset shows a schematic diagram representing the body-centred tetragonal $BaCuSi_2O_6$ crystal lattice (the lattice exhibits a subtle incommensurate distortion below 85 H, ref. 18), dimers formed from $Cu^{2+}$ $s=½$ spins are shown as dumb-bells. It is apparent that this lattice structure leads to geometrical frustration. $J$ is the intra-dimer antiferromagnetic interaction, $J'$ the antiferromagnetic interaction between in-plane dimers, and $J_f$ the inter-layer antiferromagnetic interaction that leads to geometric frustration.

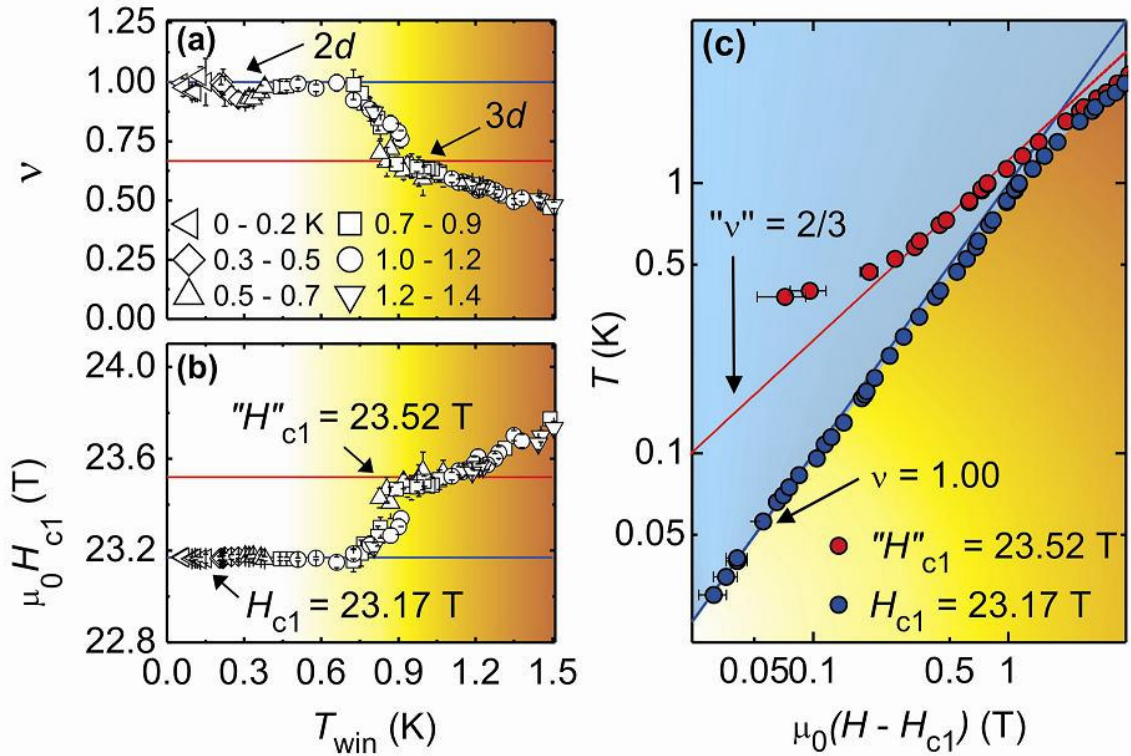

**Figure 2 | Crossover from 3D to 2D BEC critical exponent. a**, Values of the critical exponent $\nu$ obtained from fitting experimental points on the phase boundary in a sliding window centred at $T_{\text{win}}$ (K). A two-parameter least squares regression of equation (1c) with $\nu$ and critical field $H_{c1}$ varying is used to fit the data. Error bars show the standard errors in the least squares fit. The window size varies from 0.05 to 1.4 K, as indicated by different symbols. The data approach $\nu = 2/3$ in the intermediate regime, and there is a distinct crossover toward $\nu = 1$ before the QCP is reached. **b**, Estimates of $H_{c1}$ obtained along with $\nu$ during the fit. $H_{c1}$ approaches a transient value of 23.52 T in the intermediate regime, and crosses over to the true low temperature value of 23.17 T before the QCP is reached. The shading reflects the crossover toward the $\nu = 1$ exponent as the QCP is approached both as a function of field and temperature. The dark yellow shading indicates the high temperature (field) non-universal regime, the bright yellow the intermediate regime, and the light yellow the 2D regime. **c**, Best fits to the phase boundary in the intermediate and low temperature regimes represented on a logarithmic scale. The solid lines show that the data in the

intermediate temperature range are consistent with the values of $H_{c1} = 23.52$ T and $\nu = ⅔$, which does not fit the lower temperature points; whereas data in the lower temperature range are consistent with $H_{c1} = 23.17$ T and $\nu = 1$, which does not fit the higher temperature points. We observe a crossover from one regime to the other in the temperature range 0.65 K<$T_{win}$<0.9 K.

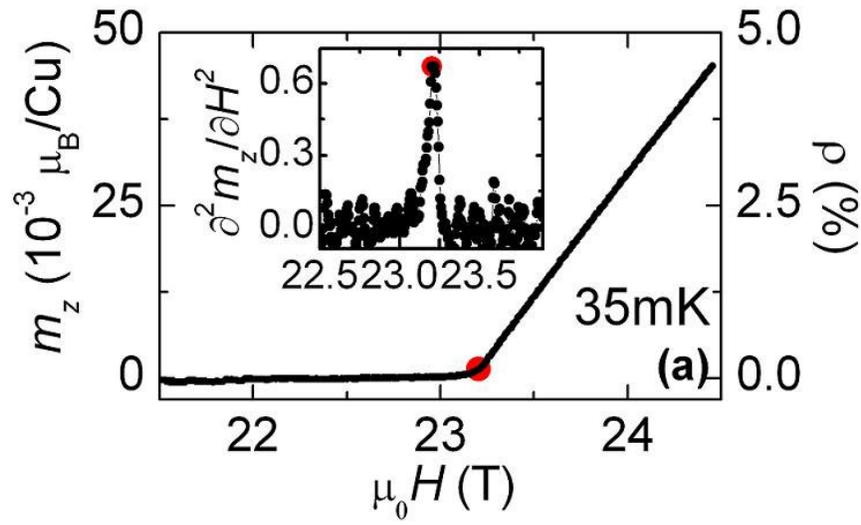
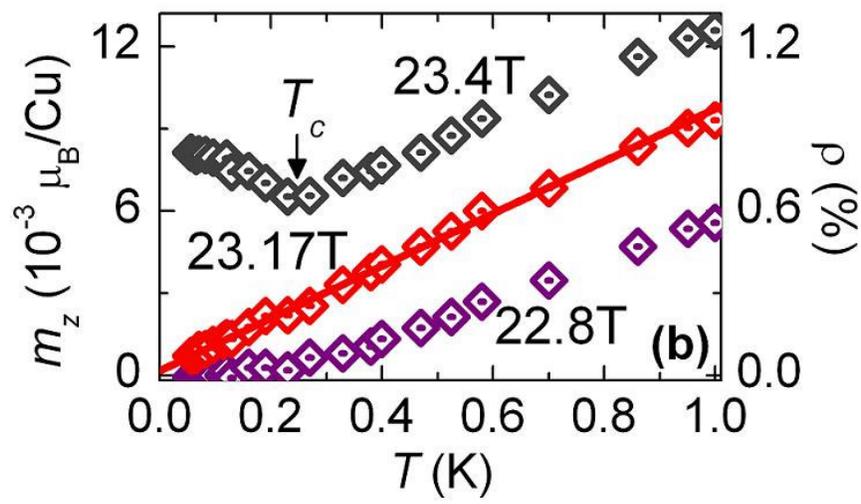
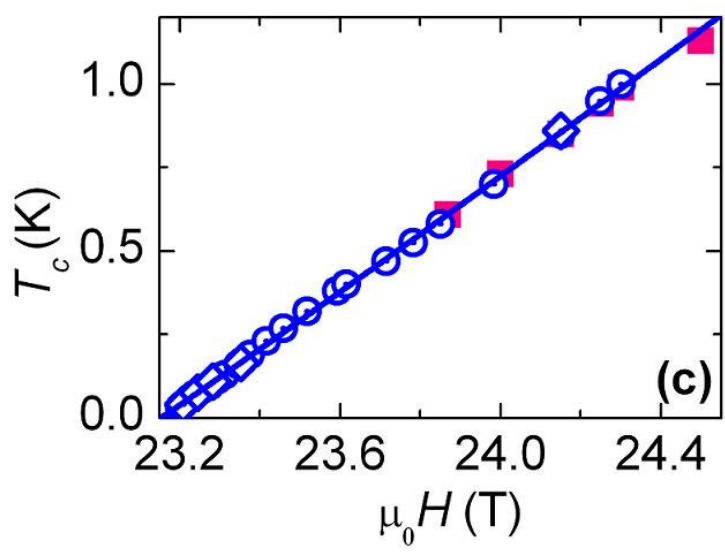

**Figure 3 | 2D BEC power law behaviour**. Power laws extracted from uniform magnetization obtained from torque measurements on $BaCuSi_2O_6$ in an external magnetic field applied at a small angle (<10°) to the crystalline *c* axis. Error bars are smaller than the symbol size. A small background subtraction has been made to account for the cantilever response; absolute values of magnetization were obtained by comparison with pulsed magnetic field data. **a**, Uniform magnetization ($m_z$) ≡ particle density ($\rho$) obtained from torque measured as a function of rising magnetic field at 35 mK. The ordering transition determined from a sharp feature in the second derivative[13] (shown in the inset) is indicated on the magnetization curve (**b**) $m_z \equiv \rho$ as a function of temperature extracted from the magnetization curves in **a** measured at various temperatures. Representative curves are shown at magnetic fields above, below and at the critical magnetic field $H_{c1}$ = 23.17 T. $T_c$ = 0 K at $H_{c1}$, and corresponds to the dip in $m_z$ at fields above $H_{c1}$. The solid line shows a linear fit to $m_z$ at $H_{c1}$. **c**, Points on the phase boundary obtained from ordering transitions determined from the magnetization curves in **a**. The solid squares represent data previously reported[13]. The open symbols represent data reported in this work which are measured at lower temperatures in a dilution refrigerator. The open circles and diamonds represent data taken with the sample in slightly different orientations (the angle was changed by ~10° and the field accordingly rescaled by a 0.5% change in *g*-factor) to verify that the results are independent of orientation. The solid line shows a linear fit to the phase boundary below 1 K.

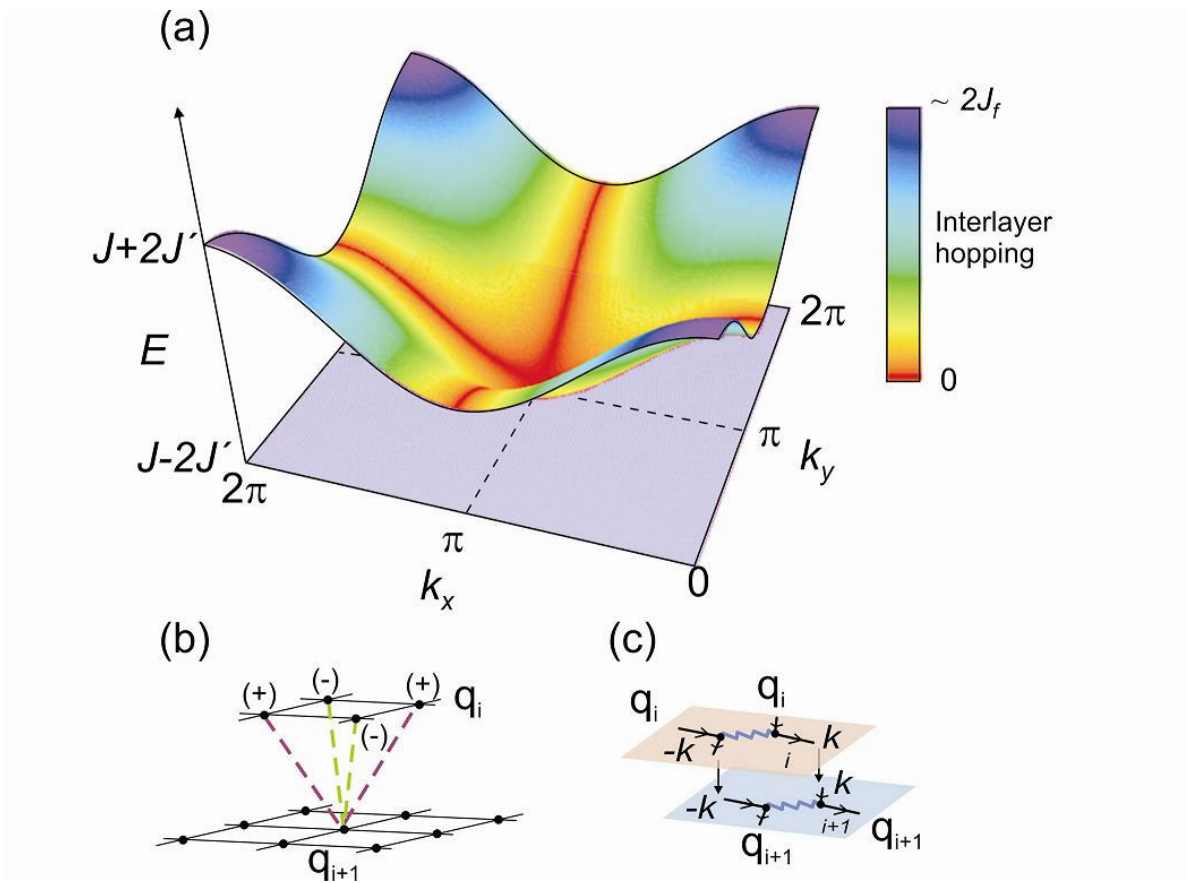

**Figure 4 | Inter-layer decoupling at the QCP. a**, Dispersing $s_z = 1$ triplon band in an applied magnetic field of $H_{c1}$. The shape of the dispersion is the same in any magnetic field $H \leq H_{c1}$, but the gap to the singlet state decreases with increasing $H$. At $H = H_{c1}$, the gap to the singlet state is closed at wavevector $k_x = k_y = \pi$, resulting in Bose–Einstein condensation. The colour shading represents the inter-layer hopping. At the wavevector where Bose–Einstein condensation occurs, there is no inter-layer hopping, which leads to a 2D BEC QCP. **b**, A diagram of the phase alternation of a single boson state with $\mathbf{q} = (\pi, \pi)$, resulting in phase cancellation on the next layer (that is, no interaction between the two layers). **c**, Inter-layer tunnelling due to zero-point fluctuations is equivalently represented as interaction-induced pair hopping, proportional to $\rho^2$ at low boson densities.